\documentstyle[aps]{revtex}

\newcommand{\beq}{\begin{equation}}
\newcommand{\beqa}{\begin{eqnarray}} 
\newcommand{\eeqa}{\end{eqnarray}} 
\newcommand{\eeq}{\end{equation}}
\newcommand{\dg}{^{\dagger}}
\def\reals{\hbox{\rm I\kern -.2em R}}
\def\sdd{Schr$\ddot{\mbox{o}}$dinger }
\def\sdds{Schr$\ddot{\mbox{o}}$dinger's }
\def\com{\hbox{\rm l\kern -.4em C}}
\def\tr{\mbox{Tr}}

\def\dfn{\hbox{\rm $=$\kern -.95em $~^{^{\triangle}}$}}

\begin{document}


\title{Entanglement of a Double Dot with a Quantum Point Contact}

\author{Michael Steiner\footnote{E-mail: mjs@mike.nrl.navy.mil} and R. W. Rendell\footnote{rendell@estd.nrl.navy.mil}}
\address{Naval Research Laboratory, 
Washington, DC 20375}
\date{\today}
\maketitle 

\begin{abstract} 
Entanglement between particle and detector is known to be inherent in the measurement process. Gurvitz recently analyzed the coupling of an electron in a double dot (DD) to a quantum point contact (QPC) detector.  In this paper we examine the dynamics of entanglement that result between the DD and QPC. The rate of entanglement is optimized as a function of coupling when the electron is initially in one of the dots. It decreases asymptotically towards zero with increased coupling. The opposite behavior is observed when the DD is initially in a superposition: the rate of entanglement increases unboundedly as the coupling is increased. The possibility that there are conditions for which measurement occurs versus entanglement is considered.
\end{abstract}


\section{Introduction}

The relationship of entanglement to the problem of measurement appears to have been first considered by \sdd \cite{Q1:Schrodinger}.  It is well known that \sdds equation predicts that the particle and measurement device become entangled \cite{Q1:Bub}[p. 33] when the particle is initially in a superposition of eigenstates of the measurement device. Additionally, decoherence is often related to entanglement as seen in an example analyzed in \cite[p. 222]{Q1:Bouwmeester}. 

Gurvitz \cite{Q1:Gurvitz1}\cite{Q1:Gurvitz2} recently has made an important contribution in the study of the measurement problem in his analysis of an electron in a double-dot (DD) interacting with a quantum point contact (QPC).  His analysis considers the two components of the system: the current through the QPC and the density matrix of the electron. 
In this paper, we extend Gurvitz's analysis by quantifying the dynamical entanglement that results between the DD and QPC.

In Sec. \ref{secback}, Entanglement and Gurvitz's model of a DD-QPC is briefly reviewed. In Sec. \ref{secent} entanglement and its rate of change is derived for the DD-QPC. 
In Sec. \ref{secddqpc} we examine the case when the electron initially starts off in the left dot. Several properties of entanglement are identified. We investigate the rate of entanglement as a function of coupling. It was found that although the optimal coupling that maximizes the rate of entanglement is in general a function of time, there is a single coupling that is nearly uniformly optimal for all time.  The maximal rate for this case is defined as the maximal natural rate of entanglement.

In Sec. \ref{secsup} the case when the electron is initially in a superposition of  right and left dots is further quantified. For this case, several more unexpected properties of entanglement are identified. We found that the entanglement is proportional to the coupling in a manner reverse to the case where the electron started off on one of the dots and naturally entangles with the QPC. The entanglement also increases at an unexpectedly faster rate compared to the case when the electron is initially in an eigenstate of the measurement device. The rate appears to be unlimited as the coupling increases.  Due to these results, several new possibilities are suggested in Sec. \ref{secco}

\section{Background}
\label{secback}

\subsection{Entanglement}

In order to examine the role of entanglement for this paper we utilize as a measure the entropy of entanglement of a pure composite bipartite state \cite{Q1:Popescu1}.  The entropy of entanglement is not a unique measure \cite{Q1:Vidal} in terms of characterizing the Schmidt coefficients of the state. It is a measure that is nonzero iff (if and only if) the state is entangled.

Let $D$ represent the state of a system and $Q$ the environment including detector.  For the purposes of this paper, $D$ is associated with the double quantum dot, and $Q$ is approximated by the quantum point contact detector.  Let $D\in K(H) $ where $K(H)$ is defined as the set of density operators, i.e. unit trace non-negative Hermitian operators on the Hilbert space $H$. Let $\rho_{1}= D(0) \otimes Q(0)$ be the initial state of particle and detector.  Assuming $\rho_{1}$ is a pure state, then both $D(0)$ and $Q(0)$ must be pure due to the tensor product form. Now, consider an interaction Hamiltonian between $D$ and $Q$ represented by a unitary evolution $U$ so that the final state is $\rho_{2}=U \rho_{1} U'$. Typically, decoherence is observed in $D$ when the environment is traced from $\rho_{2}$ as shown for example in \cite{Q1:Omnes}[Sec. 7.1]. That is, when one computes
\beq \rho_{2}^{(D)} = \tr_{Q} \rho_{1}, \eeq 
the off-diagonal elements of $\rho_{2}$ are small, and the state is in-general mixed.

The entropy of entanglement $({\cal S}) $ of a bipartite pure state $\rho\in K(H_{A} \otimes H_{B})$ is given by Von Neumann's entropy of either $\rho_{A}$ or $\rho_{B}$,
\beqa 
{\cal S}(\rho)  & =  & -\tr (\rho_{A} \log \rho_{A} ) \nonumber \\
& =  & -\tr (\rho_{B} \log \rho_{B} ) \nonumber \\
\label{eqnent} & =  & -\sum_{i} (\lambda_{i} \log \lambda_{i} ) \eeqa
where $\rho_{A}=\tr_{B} \rho$, $\rho_{B}=\tr_{A} \rho$ and $\lambda_{i}$ are the eigenvalues of either $\rho_{A}$ or $\rho_{B}$ (also one can use the square of the Schmidt coefficients in the expansion of the composite pure state as in \cite{Q1:Bennett}). 

It is at once clear that when a component  (that is a part of a larger pure state) decoheres in the sense that the reduced density matrix becomes a mixed state, then ${\cal S}(\rho)>0$. This follows from Eqn. (\ref{eqnent}) since all mixed states have at least two non-zero eigenvalues, one of which must be strictly less than unity. This confirms that as a system decoheres in a unitary process, entanglement must be present.

The results in this section apply to a bipartite pure state $\rho\in K(H_{A} \otimes H_{B})$ where the dimension of $H_{A}$ or $H_{B}$ is arbitrary. That is, ${\cal S}(\rho)>0$ when either component state is mixed and ${\cal D}(H_{A})=2$ and ${\cal D}(H_{B})>>1$.  This is important, as this will apply later for a particular model of the DD-QPC entanglement.

We measure coherence as \beq \label{eqncoh} ||\rho_{2}- \mbox{DIAG}(\rho_{2})|| \eeq where $||x||$ denotes the L2 or Euclidean norm, DIAG($A$) represents a diagonal matrix with the  diagonal elements of $A$. 

\subsection{DD-QPC}

Gurvitz \cite{Q1:Gurvitz1} \cite{Q1:Gurvitz2} considered the measurement of a single electron oscillating in a double-dot by using a quantum point contact detector.  We briefly review the setup in Fig. \ref{figgur}, but refer the reader to the papers. The barrier shown in the figure is connected with two reservoirs at the potentials $\mu_{L}$ and $\mu_{R}=\mu_{L}-V_{d}$ respectively where $V_{d}$ is the applied voltage.  The Hamiltonian $H$ consisting of the point contact Hamiltonian $H_{PC}$, double-dot $H_{DD}$ and their interaction $H_{int}$ is
\[ H= H_{QPC} + H_{DD} + H_{int} \] where 
\begin{eqnarray*}
H_{QPC} & = & \sum_{l} E_{l} a_{l}\dg a_{l} +\sum_{r} E_{r} a_{r}\dg a_{r} + \sum_{l,r} \Omega_{lr}(a_{l}\dg a_{r}+ a_{r}\dg a_{l})  \\
H_{DD} & = & E_{1} c_{1}\dg c_{1} + E_{2} c_{2}\dg c_{2} + \Omega_{0}(c_{2}\dg c_{1} + c_{1}\dg c_{2}), \\
H_{int} & = & -\sum_{l,r} \Omega_{lr}c_{2}\dg c_{2}(a_{l}\dg a_{r}+ a_{r}\dg a_{l}), 
\end{eqnarray*}
and $E_{l,r}$ are the energy levels in the respective reservoirs, $\Omega_{lr}$ is the coupling between the reservoirs, $\Omega_{0}$ is the coupling between the left and right dots. Transmission coefficients $T_{1}$ and $T_{2}$ are assumed so that the current that flows through the QPC is $I_{1}= e T_{1} V_{d}/(2\pi)$ when the electron occupies the left dot and $I_{2}=e T_{2} V_{d}/(2\pi)$ for the right dot. Gurvitz assumed without loss of generality that $T_{2}=0$.  $T_{1}$ is selected so that the difference in the currents $I_{1}-I_{2}$ is macroscopically large.
 For this problem, Gurvitz derived the following rate equations for the time-evolution of the density matrix of the DD:
\begin{eqnarray}
\label{eqnrate}
\dot{\sigma_{11}} & = & i \Omega_{0} (\sigma_{12} - \sigma_{21}) \\
\dot{\sigma_{22}} & = & i \Omega_{0} (\sigma_{21} - \sigma_{12}) \nonumber \\
\dot{\sigma_{12}} & = & i \epsilon \sigma_{12} + i \Omega_{0} (\sigma_{11} - \sigma_{22}) - \frac{1}{2} \Gamma_{d} \sigma_{12} ,\nonumber
\end{eqnarray}
where $\Gamma_{d}= T_{1} V_{d}/(2 \pi)$ and $\epsilon= E_{2}-E_{1}$.

\section{Entanglement for the DD-QPC}
\label{secent}

Suppose that the initial state of the DD is a pure state parameterized by $\theta,\phi\in[0,2\pi]$ such that $\sigma_{11}(0)= \cos^{2} \frac{\theta}{2}$, and $\sigma_{12}(0)= \sin \frac{\theta}{2} \cos \frac{\theta}{2} \exp(-i\phi)$. Then the entropy of entanglement, ${\cal S}$ is characterized by the eigenvalues of the density matrix $\sigma(t)$ of the DD. For the case of aligned levels, $\epsilon=0$, inserting the initial conditions into Eqn. (\ref{eqnrate}) and solving yields
\begin{eqnarray} 
\label{eqnsigma}
\sigma_{11}(t) & = & \frac{1}{2}[1+\cos \theta \exp(-\frac{\Gamma_{d} t}{4}) (\cosh \frac{\omega t}{4}+\frac{\Gamma_{d}}{\omega}\sinh \frac{\omega t}{4})] +  \\
&  & 4 \frac{\Omega_{0}}{\omega} \sin \theta \sin \phi \exp(-\frac{\Gamma_{d} t}{4}) \sinh \frac{\omega t}{4}  \nonumber \\
\sigma_{12}(t) & = & \frac{1}{2}\sin \theta \cos \phi \exp (-\frac{1}{2}\Gamma_{d} t) + i \exp(-\frac{\Gamma_{d} t}{4})[ \frac{4 \Omega_{0}}{w} \cos \theta \sinh \frac{\omega t}{4}  -   \\ \nonumber
& &  \frac{1}{2} \sin \theta \sin \phi (\cosh \frac{\omega t}{4} - \frac{\Gamma_{d}}{\omega} \sinh \frac{\omega t}{4})] 
\end{eqnarray}
where $w \dfn (\Gamma_{d}^{2}-64 \Omega_{0}^{2})^{\frac{1}{2}}$, $e_{\pm}\dfn \frac{1}{4} (\Gamma_{d}\pm w)$. Now, the entropy can be computed as 
\beq
\label{eqnentr}
{\cal S}(t) = - [\lambda_{+} \log \lambda_{+} + \lambda_{-} \log \lambda_{-} ] \eeq
where $\lambda_{\pm}$ are the eigenvalues of $\sigma(t)$, given by
\[ \lambda_{\pm} = \frac{1}{2} [1 \pm \sqrt{1-4(\sigma_{11}\sigma_{22}-|\sigma_{12}|^2)}], \] with $\lambda_{+}\ge\lambda_{-}$, $\lambda_{+}=1-\lambda_{-}$.

Now, the rate of entanglement, ${\cal R} \dfn \frac{d S}{d t}$ can be computed from (\ref{eqnentr}) as,
\beq \label{eqnenrate} {\cal R}(t) = \frac{\Gamma_{d} |\sigma_{12}|^{2}}{(2 \lambda_{+}-1)} \log (\frac{\lambda_{+}}{1-\lambda_{+}})  \eeq
\\

\noindent {\it Remark: The Entropy of Entanglement monotonically increases for all $t$ and for all initial conditions of $\theta,\phi$.} \\

Note that this monotonicity property is not in general guaranteed. The case of a two-level atom coupled to a cavity is well-known to result in a non-monotonic, collapse and revival, behavior of the entropy \cite{Q1:Buzek} \cite{Q1:Farhadmotamed}.

\section{Double-Dot initially in left Dot coupled to a QPC}
\label{secddqpc}

Consider the case when the double-dot is initially localized to one of the dots, let us assume the left dot. In this case $\theta=0$.  Note from Eqn (\ref{eqnenrate}) that when $\theta=0$, ${\cal R}(t)$ is continuous in $t$. We define $\alpha\dfn \Gamma_{d}/\Omega_{0}$ and $\tau\dfn \Omega_{0} t$. Hence, $\alpha$ is a normalized measure of the coupling between the DD and QPC, and $\tau$ is normalized to $\Omega_{0}$.  The case where $\alpha>>1$ is strong coupling whereas for $\alpha<<1$ is weak coupling.  Entanglement, defined by Eqn. (\ref{eqnentr}), is plotted as a function of time in Fig. \ref{figent} for  several coupling parameters $\alpha=.1,1,10,100,1000$. There are several interesting features that will be discussed.

\subsection{Entanglement and Coherence}

Note in Fig. \ref{figent}, for weak damping, that the entanglement often increases and then stops, and then later starts to increase again. It is seen from Eqn. (\ref{eqnenrate}) that when $\sigma_{12}$ is zero, then the rate of entanglement is zero. In Fig. \ref{figentcoh} we overlay the entanglement (labeled with `S') and coherence (labeled with `C') as defined by Eqn. (\ref{eqncoh}) for the cases of $\alpha=.01$ and $\alpha=1$. Note that  it appears that the coherence terms is cycling in time roughly on the order of $\tau=1.6$ for weak enough damping. For each cycle, the entanglement increases rapidly when the coherence is maximal and stops increasing when the coherence is zero.  For the case of $\theta=0$, Eqn. (\ref{eqnsigma}) can be rewritten in the form 
\[ |\sigma_{12}(\tau)|^{2} = \frac{16}{\alpha^{2}-64} \exp{-\frac{\alpha \tau}{2}} \sinh^{2} (\frac{\tau}{4}\sqrt {\alpha^{2}-64}  ). \] 
Note that $\sigma_{12}$ will  oscillate when there is a complex term in the Sinh function. No cycling was found for the case of strong damping when $\alpha>8$. The cycling time was found to lengthen as $\alpha<8$ approaches $8$. 

\subsection{Rate of Entanglement}

The rate of entanglement for the case $\theta=0$ is bounded both for all time and for all $\alpha$. For small $\tau$, ${\cal R}(\tau)=0$ at $\tau=0$. Additionally, it is found for small $\tau$ that ${\cal R}(\tau)$ is inversely proportional to the coupling for weak damping.  It is found that $R(\tau)\rightarrow 0$ with $\tau$ and also with increasing coupling for large $\tau$.  It is found that $R(\tau)\rightarrow 0$ as $\tau\rightarrow \infty$ and also decreases with increasing coupling for large $\tau$. 

\subsection{Existence of Optimal Coupling}

 Note from Fig. \ref{figent} that there appears to be an optimal coupling that maximizes the rate of entanglement.  This can be found by examining 
\beq \label{eqnentalpha}  \frac{d {\cal S}}{d\alpha} = \frac{1}{2\lambda_{+}-1} \log {\frac{\lambda_{+}}{1-\lambda_{+}}} \frac{d}{d\alpha}(\sigma_{11}\sigma_{22}-|\sigma_{12}|^2). \eeq
It is found that (\ref{eqnentalpha}) is not uniformly optimized at a single $\alpha$ for all $\tau$, although $\alpha=5$ is a good approximation for the coupling that maximizes the rate.  For this case, the DD becomes nearly completely entangled with the QPC on the order of unity normalized time, i.e. $\tau=1$.  Note that a similar result of the existence of an optimal coupling has been shown for a different problem in \cite{Q1:Bellnf}.

\section{Double-Dot initially in superposition coupled to a QPC}
\label{secsup}

In this section we consider from an entanglement perspective what happens when the DD is initially in a superposition state of left and right dot. In this case $\theta \ne 0$. Note from Eqn (\ref{eqnenrate}) that unlike in the previous section, ${\cal R}(\tau)$ is discontinuous at $\tau=0$ because $\lambda_{+}(\tau) \rightarrow 1$ as $\tau \rightarrow 0$.  However, it is not clear that this discontinuity is not a facet of the idealized model that is proposed. This is further discussed in Sec. \ref{secco}.  Entanglement of Eqn. (\ref{eqnentr}) is plotted as a function of time in Fig. \ref{figentsup} for  several coupling parameters $\alpha=.1,1,10,100,1000$ when $\theta=90$. There are again several new interesting features.

\subsection{Entanglement and Coherence}

Entanglement is again predicted by \sdds equation when the DD is initially in a superposition. The rise of entanglement (labeled with `S') was accompanied by a loss of coherence (labeled with `C') as is shown in Fig. \ref{figentsupcoh}. However, we did not see the same cycling effect that was seen in Fig. \ref{figentcoh} for the case when the particle was localized to the left dot.

\subsection{Rate of Entanglement}
 
For the case of $\theta \ne 0$, the rate of entanglement of Eqn. (\ref{eqnenrate}) is for small $\tau$ dominated by,
\beq \label{eqnsmtau} {\cal R} \propto \alpha \log (\frac{\lambda_{+}(\tau)}{1-\lambda_{+}(\tau)}) \exp(-\frac { \alpha \tau}{2}) \eeq 
where $\lambda_{+}(\tau)\rightarrow 1$ as $\tau\rightarrow 0$, and for large $\alpha$
\beq {\cal R} \propto \exp(-\frac{\alpha \tau}{2}). \eeq
The rate of entanglement generally decreases in time. This is in contrast to the case of $\theta=0$ where the rate of entanglement initially increases from zero (because $|\sigma_{12}(0)|^{2}=0$), reaches a maximum, and then asymptotically decreases.  

\subsection{Dependencies on Coupling Parameters}

The most significant differences that were seen between this case and the case when the DD is initially in a superposition are on the dependencies of the coupling parameters. As can be seen from Fig. \ref{figentsup}, the entanglement increases very rapidly initially, particularly as the coupling is increased.  In contrast to the case when the DD is initially localized, there is no optimal coupling parameter that optimizes the rate of entanglement. In fact, the rate of entanglement appears to be unbounded with the coupling. Hence the great difference between the two cases of localized electron versus superposition is apparent when the coupling is large; in the case of localized electron, the entanglement will take longer with increased coupling; in the case of a superposition, the entanglement will be shorter with increased coupling. Hence there are substantial differences seen in the \sdd predicted entanglement for the cases of $\theta=0$ and $\theta=90$. 

For the case of $0<\theta<90$, one sees for small $\tau$ a similar dependence as is given for $\theta=90$, while for longer $\tau$ the dependence shifts to what is seen for $\theta=0$. An example is shown in Fig. \ref{figentpsup} for $\theta=30$.

\section{Conclusions}
\label{secco}

The process of measurement under \sdd evolution will enjoin entanglement between particle and measurement device. The recent analysis by Gurvitz has been examined in terms of quantifying the entanglement between the DD and QPC. A number of features related to entanglement have been shown. 

It is not clear if  features such as the discontinuity seen in ${\cal R}$ for the case of $\theta\ne 0$ are related to overly-idealized modeling of the QPC. An area of refinement that is most likely needed is in the model of the QPC as a pure state. In reality, measuring devices are in a mixed state due to thermal effects as noted for example in \cite{Q1:Bose}. Such an entanglement analysis would be more complicated but is necessary for a more complete analysis.

When \sdds equation predicts entanglement, under what conditions will a measurement occur? Although we cannot answer this question definitively at this time, it appears reasonable to pose the question. As we have seen, the rate of entanglement is substantially different for the case where the electron is initially in the left or right dot versus when it is in a superposition. Entanglement may occur for the former case where the entanglement rate is bounded, whereas the rate is unbounded for the latter case and this is precisely the case when measurement is classically expected.   However, more evidence and refinements will be needed to answer the question posed.

\vspace{.2in}

\noindent {\bf ACKNOWLEDGEMENTS}
The authors would like to thank Dr. Peter Reynolds for his support from the Office of Naval Research.


\newpage

\begin{figure}
\caption{Double Dot and Quantum Point Contact}
\label{figgur}
\end{figure}

\begin{figure}
\caption{Entropy versus time for various $\alpha$ when $\theta=\phi=0$}
\label{figent}
\end{figure}

\begin{figure}
\caption{Entropy (S) and Coherence (C) versus time for various $\alpha$ when $\theta=\phi=0$}
\label{figentcoh}
\end{figure}

\begin{figure}

\caption{Entropy versus time for various $\alpha$ when $\theta=90$, $\phi=0$}
\label{figentsup}
\end{figure}

\begin{figure}

\caption{Entropy (S) and Coherence (C) as a function of time for various $\alpha$ when $\theta=90$, $\phi=0$}
\label{figentsupcoh}
\end{figure}

\begin{figure}

\caption{Entropy for various $\alpha$ when $\theta=30$, $\phi=0$}
\label{figentpsup}
\end{figure}

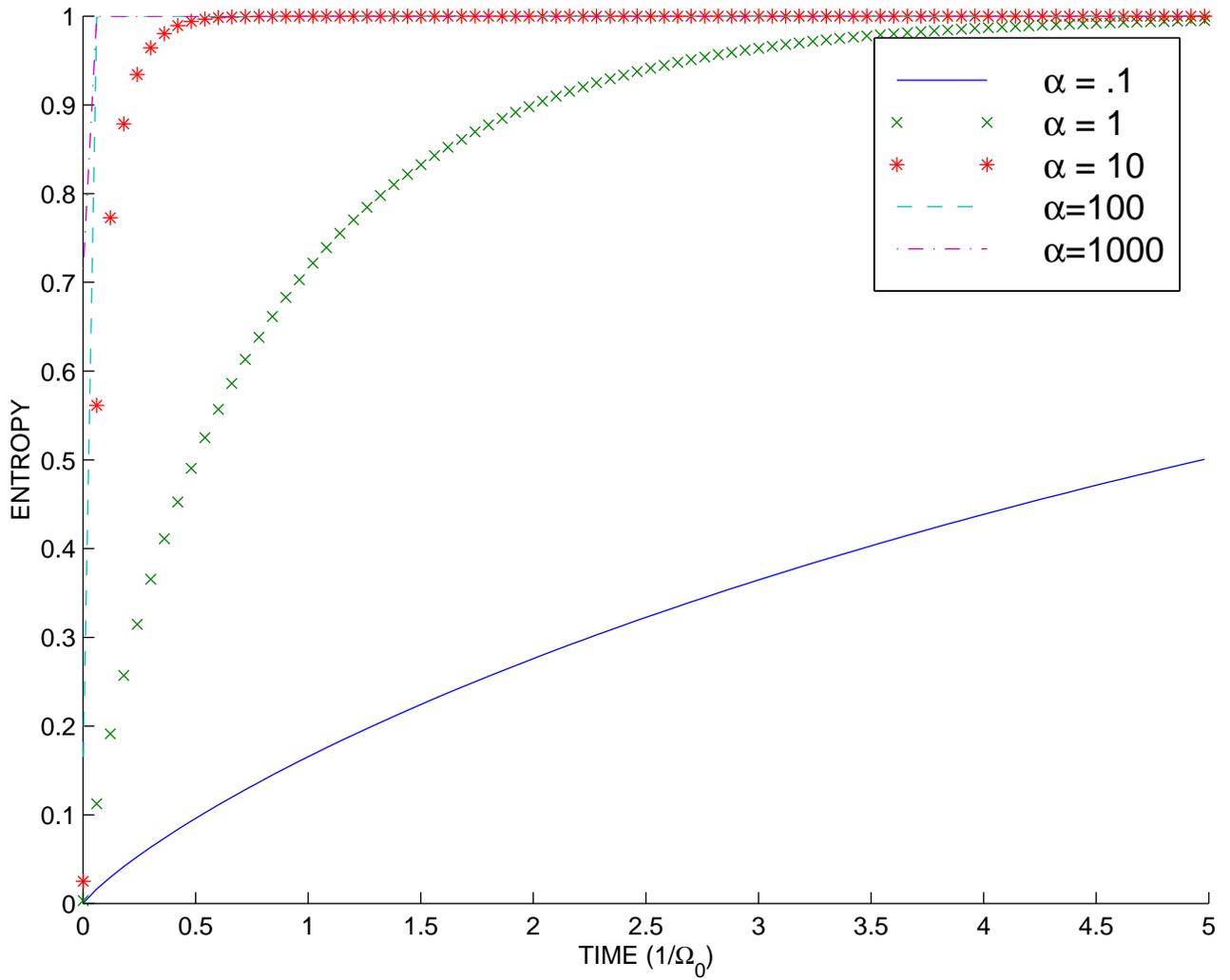
\end{document}